\documentclass[twocolumn,showpacs,showkeys,preprintnumbers,amsmath,amssymb,aps pra,superscriptaddress]{revtex4}

\usepackage{graphicx}
\usepackage{color}
\usepackage{subfigure}

\usepackage{dcolumn}% Align table columns on decimal point
\usepackage{bm}% bold math

\newcommand{\parenth}[1]{\ensuremath{\left(#1\right)}}

\newcommand{\logneg}{\ensuremath{E_{\mathcal{N}}}}

\newcommand{\W}{\hphantom{-}}
\newcommand{\jf}[1]{{\color{black} #1}}

\begin{document}

%\preprint{FINAL VERSION}

\title{Experimental characterization of Gaussian quantum communication channels}% Force line breaks with \\

\author{James DiGuglielmo}
 \affiliation{%
Institut f\"{u}r Gravitationsphysik, Leibniz Universit\"{a}t Hannover
 and Max-Planck-Institut f\"{u}r Gravitationsphysik (Albert-Einstein-Institute),\\
Callinstrasse 38, 30167 Hannover, Germany
}%
 \email{James.DiGuglielmo@aei.mpg.de}
\author{Boris Hage}
 \affiliation{%
Institut f\"{u}r Gravitationsphysik, Leibniz Universit\"{a}t Hannover
 and Max-Planck-Institut f\"{u}r Gravitationsphysik (Albert-Einstein-Institute),\\
Callinstrasse 38, 30167 Hannover, Germany
}%
\author{Alexander Franzen}
 \affiliation{%
Institut f\"{u}r Gravitationsphysik, Leibniz Universit\"{a}t Hannover
 and Max-Planck-Institut f\"{u}r Gravitationsphysik (Albert-Einstein-Institute),\\
Callinstrasse 38, 30167 Hannover, Germany
}%
\author{Jarom\'{i}r Fiur\'{a}\v{s}ek}
\affiliation{
Department of Optics, Palacky University, 17. listopadu 50, 77200
Olomouc, Czech Republic
}%
\author{Roman Schnabel}
 \affiliation{%
Institut f\"{u}r Gravitationsphysik, Leibniz Universit\"{a}t Hannover
 and Max-Planck-Institut f\"{u}r Gravitationsphysik (Albert-Einstein-Institute),\\
Callinstrasse 38, 30167 Hannover, Germany
}%

\date{May 5, 2007}

\begin{abstract}
We present a full experimental characterization of continuous
variable quantum communication channels established by shared
entanglement together with local operations and classical
communication.  The \jf{resulting teleportation} channel was fully characterized
by measuring \textit{all} elements \jf{of the covariance matrix} of the
shared two-mode squeezed Gaussian state.  
\jf{From the experimental data we determined}
the lower bound to the quantum channel capacity, the teleportation
fidelity of coherent states, \jf{and} the logarithmic negativity and the
purity \jf{of the shared state}.  Additionally, a positive secret key rate was obtained for two of the
established channels.
\end{abstract}

\pacs{03.67.-a,03.67.Mn,03.67.Hk}
                             
\keywords{Covariance matrix, Gaussian channels, quantum channel capacity}%Use showkeys class option if keyword
                              %display desired
\maketitle

\section{Introduction}\label{intro}

Continuous variable quantum communication channels have been the subject of both
theoretical and experimental research for the past few years
\cite{caves1994,cerf2005,eisert2005,giovannetti2003,hausladen1996,schumacher1997,holevo2001,wolf2006,grangier2004,grangier2003,schnabel2003}.
 Similar to classical communication channels, quantum communication
 channels are characterized by a channel capacity.  In
contrast to classical communication channels, the capacity of quantum
communication channels is distinguished by two different quantities;
namely, the \textit{classical capacity} \jf{which gives the number of
classical bits that can be faithfully transmitted per use of the channel} 
and the \textit{quantum capacity} \jf{which specifies how many quantum bits 
can be transmitted per use of the channel} \cite{giovannetti2003,nielsen2000}.  One example of a quantum
communication channel is a teleportation channel, which is established
by a shared entangled state with local operations and classical
communication (LOCC) between two distant parties
\cite{braunstein1998, bowen2003}.  Of all the possible entangled states that
could be used to establish the quantum channel, \textit{Gaussian}
states are of particular interest due to their well understood
theoretical structure and ability to be easily generated
experimentally \cite{eisert2005,wolf2006}.  Because these states are characterized by a
Gaussian \jf{Wigner} function, only the second moments collected in
the state's covariance matrix (CM) are required in order to completely define the state.  Experimentally,
this means that only a few tomographic measurements need to be
conducted, significantly reducing the effort to measure these states.
To date, several groups have conducted experiments only
\textit{partially} measuring the CM \cite{grangier2005,bowen2004,laurat2005}.  

This paper presents an experimental study of Gaussian
quantum teleportation channels.  The teleportation channels are
established by distributing two different classes of entangled
Gaussian states \jf{illustrated in Fig.~\ref{ent_class}} over a free-space auxiliary channel to two parties,
Alice and Bob, together with local operations and classical
communication (LOCC).  In \jf{our experiment}, every single parameter of
the CMs are measured.  These channels are \jf{then} characterized by evaluating 
the lower bounds to the quantum channel capacity, the teleportation fidelities 
of coherent states, and the purities and the logarithmic negativities 
\jf{of the shared entangled states}.  Additionally, two different entanglement
criteria are used-the Simon-Peres-Horodecki and an entanglement
witness-to verify that the measured state is in fact entangled.  This paper is
divided into the following sections.  In Sec.~\ref{exp_proced}, we
present an efficient experimental procedure for measuring the entire
CM using only five measurement settings.  \jf{The technical details of our
experiment are described in Sec.~\ref{exp_setup}. The
experimental implementation of the measurement of the entire covariance matrix
is discussed in Sec. \ref{proto_imp}.}
The formal definitions of a
quantum channel as well as the quantities that characterize them are
presented in Sec.~\ref{channels}.   The reconstructed
CMs from the experimental data are presented in Sec.~\ref{results} and
finally Sec.~\ref{discussion} contains a discussion of the results.          

\section{Experimental modus operandi}\label{exp_proced}

\subsection{Preliminary Considerations}\label{pre_conds}

In order to obtain complete knowledge of a \jf{two-mode} Gaussian entangled state, it is
sufficient to measure its symmetric \jf{positive} semi-definite 
ten parameter covariance matrix (CM)
\cite{bowen2004,laurat2005}.  In its block form, the CM is given by
\begin{equation}\label{cvmat}
\jf{\gamma=}
\parenth{\begin{array}{cc}
{\bf A}&{\bf C}\\
{\bf C}^{T}&{\bf B}
\end{array}},
\end{equation}
where ${\bf A}$, ${\bf B}$ and ${ \bf C}$ are $2 \times 2$ \jf{matrices} 
which contain the parameters describing
Alice's mode, Bob's mode and the correlations between their modes,
respectively.  The CM contains the second moments of a state's
quadratures, \jf{$\gamma_{jk}=\langle \Delta r_j \Delta r_k +\Delta r_k \Delta
r_j \rangle$,}
where $r=\parenth{x_A,\,p_A,\,x_B,\,p_B}$ is a vector of quadrature
operators and $\Delta r_j = r_j - \langle r_j \rangle$.  \jf{We use units such that 
the covariance matrix of vacuum is equal to the identity matrix.}
From \jf{$\gamma$} 
 can be obtained information regarding 
entanglement properties \jf{of the state} (e.g. verification, quantification) 
as well as the state's purity.  In the case of teleportation channels, the lower
bound to the quantum channel \jf{capacity and} the
teleportation fidelity of coherent states \jf{can}
also be obtained from the CM.

\jf{For applications such as monitoring of quantum communication channels 
it is highly desirable} to develop techniques
such that the reconstruction of a state's CM can be
accomplished with the fewest possible measurements.  To this end, the
\jf{structure} of the matrix itself can be exploited such that only two
measurement settings yield six of the ten independent parameters (simultaneous 
measurement \jf{of the amplitude quadrature of one mode and the phase
quadrature of the other mode}).  Besides these more
\jf{technical} considerations regarding an efficient experimental
procedure for the detection and quantification of entangled Gaussian
states, there are a number of fundamental issues that must be addressed.
These have been elaborated upon by van Enk \textit{et
  al}.~\cite{vanenk2006} \jf{who} gave five criteria that should be obeyed
when conducting an entanglement experiment.  The heart of the criteria
is not to assume too much as to \jf{the} form, symmetry or
repeatability of the entanglement source for each copy that it
produces.   The effect of not satisfying these criteria is to increase the risk of
overestimating/underestimating the amount of entanglement present in the generated
state.  Any entanglement verification protocol should satisfy these
five criteria.  The choice of a verification protocol will ultimately depend on the
\textit{type} of entanglement generated (or thought to have been
generated) in an experiment.  

\begin{figure}
\begin{center}
\includegraphics[width = 9cm]{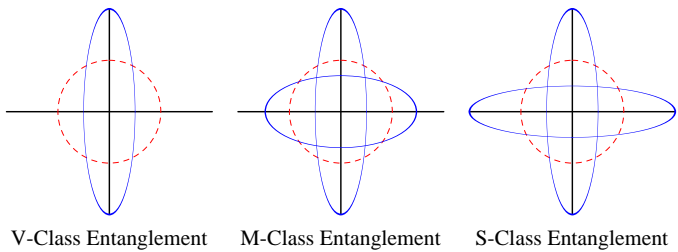}
\caption{(Color online) Classes of entanglement: This figure depicts
  three different \textit{classes} of entanglement.  V-Class
  entanglement is formed by mixing a single-mode squeezed state with
  the vacuum mode on a balanced beamsplitter (BBS). M-Class entanglement is formed by mixing two
  unevenly and oppositely single-mode squeezed beams on a BBS and S-Class entanglement is
  formed by mixing two equally but oppositely single-mode squeezed
  beams on a BBS.}
\label{ent_class}
\end{center}
\end{figure}

The establishment of a quantum communication channel, such as a teleportation channel, requires the
distribution of what van Enk \textit{et al}.~have referred to as \textit{a priori} entanglement \cite{vanenk2006}.
This type of entanglement is obtained when a source generates many
copies of a bipartite state, $\rho_{AB}$, such that an entanglement verification
protocol can be conducted on a sub-ensemble of them using the rest to
perform a quantum information theoretic protocol.  A possible
verification protocol for \textit{a priori} entanglement is to perform
full tomography on the state.  This can be achieved using linear
optics and homodyne detection \cite{breitenbach1997,leonhardt1997}.  This allows not only for a qualitative statement
as to whether the state is separable or entangled but also a
quantitative statement as to how much.  Full tomography is expensive,
however, especially when its implementation is solely to obtain
information about the channel.  As such, it is desirable to develop verification protocols
that can be conducted using only \textit{partial} tomographic measurements while
still satisfying the van Enk criteria.  We now present such a partial
tomographic protocol (PTP). 

\subsection{Description of the partial tomographic protocol}\label{direc_protocol}
The partial tomographic protocol (PTP) developed to characterize our
teleportation channels can be stated as follows:
\begin{enumerate}
\item Alice and Bob simultaneously measure their amplitude and phase
quadratures, respectively, while comparing their results by means of classical communication.
\item Alice and Bob simultaneously measure their phase and amplitude
quadratures, respectively, while comparing their results by means of classical communication.
\item Alice and Bob measure their amplitude quadratures.
\item Alice and Bob measure their phase quadratures.
\item Alice and Bob simultaneously \jf{measure} a linear combination of
their amplitude and phase quadratures, respectively.
\end{enumerate}

The fact that every parameter of the CM is measured prevents one from
making an assumption as to the symmetry of the state being measured.
Although measuring only the second moments of the state does not contain
information as to whether the state is Gaussian or not, something which
in the strictest sense of the van Enk criteria should not be
assumed, an entanglement criterion, such as the Simon criterion, is a
sufficient criterion for both Gaussian and non-Gaussian states.
Furthermore, \jf{the quantities} such as the secret key rate
\cite{garcia2006,navascues2006} or the lower bound to the quantum
channel capacity \cite{wolf2006}, while indirectly indicating the presence of
entanglement, obtain their lower bounds for Gaussian states.  As such,
one can at worst only underestimate these quantities by measuring just
the second moments and assuming that the state is Gaussian.    

\section{Experimental setup}\label{exp_setup}
\begin{figure}
\begin{center}
\includegraphics[width=9cm]{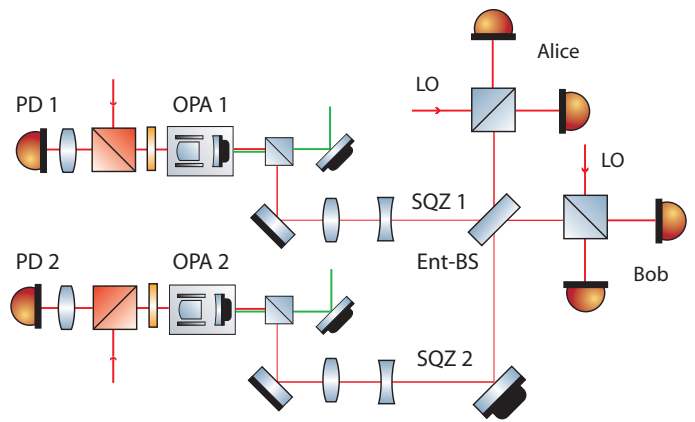}
\caption{(Color online) Experimental setup: The squeezed states are generated by the
  two optical parametric amplifiers (OPAs) and mixed at the
  entangling beamsplitter (Ent-BS).  The different classes of
  entanglement are generated by adjusting the parametric gain setting
  of the OPAs.  The entanglement is then distributed over a free-space
  channel to the two homodyne detectors Alice and Bob.}
\label{setup_pic}
\end{center}
\end{figure}

In our experiment, we generate the two-mode entangled states by mixing
on a balanced beamsplitter two squeezed vacuum beams produced by our
optical parametric amplifiers (OPAs).  The laser source used in our
experiment was a continuous-wave non-planar Nd:YAG ring laser
with 300$\,$mW of output power at 1064$\,$nm and 800$\,$mW at
532$\,$nm.  The latter was used to pump the OPAs to produce two amplitude squeezed
light beams with an approximate power of 0.06$\,$mW at 1064$\,$nm.
Both OPAs were constructed from type I non-critically phase-matched
$\mathrm{MgO:LiNbO_{3}}$ crystals inside hemilithic cavities.  Each
cavity was formed by a HR-coated crystal surface with a reflectivity
of $r > 0.999$ and a metal spacer mounted out-coupling mirror with a
reflectivity of $r=0.957$.  The intra-cavity crystal surface was AR
coated for both the fundamental
$\parenth{\mathrm{1064\,nm\,,r<0.05\%}}$ and the second harmonic
$\parenth{\mathrm{532\,nm\,,r<0.5\%}}$.  The out-coupling mirror had a
reflectivity of r=0.15$\pm$0.02 for 532$\,$nm.  The OPAs were seeded
through the HR-surface with a coherent laser beam of 15$\,$mW power
and pumped through the out-coupling mirror with various intensities,
the lowest being 75$\,$mW, corresponding to a parametric gain 5.  The
length of both OPA cavities as well as the phase of the second
harmonic were controlled using radio-frequency modulation/demodulation
techniques.  The error-signals were derived from the seed fields
reflected from the OPA cavities.  A maximum value of \jf{$4.0\,$dB} of
non-classical noise suppression was directly observed using homodyne
detection.  The shot noise level was defined by  mixing the local
oscillator with the vacuum mode on a balanced \jf{beam splitter and measuring
fluctuations of vacuum}. The
electronic dark noise of the homodyne detectors was
approximately $13\,$dB below the shot noise
level making dark noise correction of the observed squeezing
superfluous.  The visibility on both homodyne detectors was
$\eta_{\mathrm{vis}}=0.965$ and the quantum efficiency of the
photodetectors is estimated to be $\eta_{\mathrm{quantum}}=0.93$
yielding a total detection efficiency of $\eta \approx 0.87$.  The
phase locks on both the entangling beampslitter and homodyne
beamsplitters are estimated to be within $3^{\circ}$ of the desired
values.  The photocurrents produced from the homodyne detectors were
first demodulated at a frequency of 7$\,$MHz and low-pass filtered
with a corner frequency of 30$\,$kHz.  It was sampled with a National
Instruments sampling card with maximum sampling rate of 1 mega sample
per second.  By independently changing the parametric gain of each
amplifier we can generate all three types of entanglement as
illustrated in Fig.~\ref{ent_class}.  A diagram of the full experiment is provided in Fig.~\ref{setup_pic}.

\section{Experimental implementation of PTP}\label{proto_imp}
Our partial transposition protocol (PTP) was implemented using a custom built data acquisition system
whose software component was developed using LabView and whose
hardware component was realized by balanced homodyne detection
with external addition and subtraction boxes.  The homodyne detectors were designed such that there were multiple
outputs of both the DC and AC signals generated by each detector.
The AC-subtracted signals from both homodyne detectors were fed
simultaneously into the LabView program where both the variances of
the respective electronic channels as well as the covariance of the
two electronic channels were calculated in real-time.  This corresponds to the classical
communication component of our protocol.  Additional information as to which
quadratures produced a given covariance was obtained by recording, in
real-time, the DC substracted signal from one scanned homodyne
detector.  This was achieved by locking e.g.~Alice's homodyne detector to one quadrature and scanning the
phase between the local oscillator and signal beam of Bob's homodyne
detector.  With this setup, one measurement-e.g.~measuring the
amplitude quadratures simultaneously-delivered three of the ten
required CM parameters.  

There are two main features of our implementation that are noteworthy.
First, it allows the manual setting of the measurement basis.  The
basis information can be obtained by looking at the covariance of the
two electronic channels.  A zero covariance indicates the measurement
of two orthogonal quadratures for a symmetric state.  Although both the verification and
quantification of entanglement is basis independent, the form of the
CM is not.  For the case of an optimally entangled EPR state, one
would expect non-zero parameters for half the elements of the CM in an
orthogonal measurement basis.  While the choice of a basis is
arbitrary, it must be consistent.  Failure to measure every parameter
of the CM in the same basis is tantamount to random experimental
error.  Failure to measure some of the parameters in the same basis is
systematic error, as it adds a constant offset to only some of the
parameters.  \jf{The implication of these error sources, especially
systematic error, is to obtain a false estimate of quantities of
interest such as the logarithmic negativity \cite{vidal2001}, or to
reconstruct a non-physical state.} 

Second, systematic error can be reduced.  As a result of the real-time evaluation of
the covariance between the homodyne detector outputs and the DC
subtracted signal from a scanned homodyne detector, one can determine
which quadratures are correlated, anti-correlated and not correlated.
This information helps to reduce the systematic error because it
provides a means by which to \jf{adjust} the phase angle between the
optical local oscillator and signal beam independent of any DC offsets
on the \jf{error}-signal.  This contributes to the overall consistency of
the entanglement detection.  
  
\section{Theoretical description of quantum communication channels}\label{channels}
In order to more deeply understand the equivalence between a shared entangled
state and an established quantum communication channel, such as for a
teleportation channel, it is necessary to understand the theorectical structure of quantum
communication channels.  It is also within this framework that these
channels obtain their physical meaning.  To this end, this section
will review the necessary theoretical concepts in order to understand
the experimental results of Sec.~\ref{results}.  

A quantum channel is a trace-preserving completely positive map, $T$, that transforms
quantum states according to $\rho \mapsto T \parenth{\rho}$ \cite{holevo2001}.  
They \jf{can be} understood to originate as the result of a unitary interaction $U$ of a state, $\rho$, with the
environment described by another Hilbert space $\mathcal{H}_E$ which
is in a state, $\rho_E$,
\begin{equation}
T\parenth{\rho}=\mathrm{Tr}_E U\parenth{\rho\otimes\rho_E}\jf{U^{\dagger}},
\end{equation}
where $\mathrm{Tr}_E$ denotes the partial trace with respect to
$\mathcal{H}_E$ \cite{holevo2001,eisert2005,leuchs2005}.  An important subclass of these channels are
\emph{Gaussian channels}, which are characterized by a Gaussian
unitary $U$, determined by a quadratic bosonic Hamilitonian, and a
Gaussian state $\rho_E$ \cite{eisert2005}.  At the level of covariance matrices (CMs),
\jf{which offer a complete description of Gaussian states} and would 
be measured in all practical applications of \jf{continuous-variable}
quantum information protocols, the action of a channel is given by
\begin{equation}\label{cm-channel}
\gamma \mapsto X^{T}\gamma X + Y.
\end{equation}
The condition to ensure that the transformation is completely positive
is given by
\begin{equation}\label{cp-cond}
Y+i\Omega-iX^{T}\Omega X \geq 0,
\end{equation}
where
$$\Omega=\parenth{
\begin{array}{cc}
\sigma&0\\
0&\sigma\end{array}}
$$
is the symplectic form with 
$$\sigma = \parenth{
\begin{array}{cc}
0&1\\
-1&0\end{array}}.
$$
The formula Eq.~\parenth{\ref{cp-cond}} represents the necessary and
sufficient condition for complete positivity of the Gaussian map given
by Eq.~\parenth{\ref{cm-channel}} of the manuscript, see e.~g.~
Refs.\cite{demoen1977,lindblad2000,eisert2002,fiurasek2002,jamiolkowski1972}.
It is indeed possible to interpret this condition as the generalized
Heisenberg inequality.  According to the Jamiolkowski isomorphism
\cite{jamiolkowski1972}, every completely positive map is isomorphic
to a positive semidefinite operator on the tensor product of Hilbert
spaces of input and output states.  In the case of Gaussian CP maps
this operator becomes and infinitely squeezed Gaussian state
characterized by matrices X and Y.  The generalized Heisenberg
inequality for the covariance matrix of this state is equivalent to
Eq.~\parenth{\ref{cp-cond}}, c.~f.~ Ref.~\cite{fiurasek2002}

\begin{table*}
\caption{\label{sum_results} This table summarizes the channel
  characteristics for each class of entanglement used to establish a
  teleportation channel. Beginning with the first, they include: the
  state condition $\lambda$, the Simon criterion $\lambda^{T_A}$, the optimal entanglement
  witness $\mathcal{W}$, the log-negativity \logneg, the lower bound to the quantum channel
  capacity $Q_L$, the teleportation fidelity of coherent states
  $\mathcal{F}$, the purity $\mu$ and the secret key rate $K$.}  
\begin{ruledtabular}
\begin{tabular}{ccccc}
 &\multicolumn{2}{c}{V-Class}&\multicolumn{2}{c}{S-Class}\\
 Characteristic &  Gain 5&             Gain 10 &          Gain 5& Gain 10\\ 
\hline
$\lambda$        &\W0.033$\pm0.004$ &\W0.034$\pm0.003$ &\W0.063$\pm0.003$ &\W0.175$\pm0.005$ \\
$\lambda^{T_{A}}$&-0.317$\pm0.004$&-0.349$\pm0.003$&-0.600$\pm0.001$  &-0.566$\pm0.004$\\
 $\mathcal{W}$  &-0.341$\pm0.004$&-0.383$\pm0.003$&-0.599$\pm0.001$&-0.566$\pm0.004$\\
 \logneg        &\W0.602$\pm0.003$ &\W0.700$\pm0.004$ &\W1.342$\pm0.005$ &\W1.331$\pm0.009$ \\
 $Q_L$            &-0.071$\pm0.003$&-0.059$\pm0.004$&\W0.387$\pm0.005$ &\W0.100$\pm0.009$\\
 $\mathcal{F}$  &\W0.586$\pm0.003$ &\W0.597$\pm0.003$&\W0.701$\pm0.003$&\W0.695$\pm0.005$ \\
$\mu$           &\W0.648$\pm0.002$ &\W0.563$\pm0.001$ &\W0.608$\pm0.002$ &\W0.301$\pm0.002$\\
$K$             &                  &                  &\W0.323$\pm0.005$&\W0.120$\pm0.006$
\end{tabular}
\end{ruledtabular}
\end{table*}
The usual quantum information protocols\jf{, e.g.~teleportation and
quantum memory,} can all be considered as quantum channels \cite{wolf2006}.  In
this paper, we consider a special subclass of teleportation channels
established by means of a shared entangled state together with local
operations and classical communication (LOCC).  An important
characteristic of teleportation channels, as well as quantum channels
in general, is their capacity to transmit \emph{quantum
  information}, quantified in units of qubits.  To this end the
\emph{quantum capacity}
\cite{eisert2005,shor2002,devetak2005,lloyd1997} of an arbitrary
channel, $T$, is given by
\begin{align}
Q\parenth{T}&=\lim_{n \rightarrow \infty}\frac{1}{n}\jf{\sup_\rho} J\parenth{\rho\,,T^{\otimes
    n}},\label{qcap}\\
J\parenth{\rho,T}&=S\parenth{T\parenth{\rho}}-S\parenth{\parenth{T\otimes
\jf{\mathrm{id}}}\parenth{\psi}},\label{coherent_info}
\end{align}
where $\psi$ is a purification of $\rho$ and $J$ is known as the
\emph{coherent information}.  The coherent information was first
introduced by Schumacher and Nielsen in connection with error
correction \cite{schumacher1996}.  With regard to its operational interpretation,
the coherent information quantifies the amount of information the
environment has obtained about the state transversing it.   Another
information theoretic quantity related to the
coherent information is the \emph{quantum conditional entropy}
\cite{slepian1972,mhorodecki2005,mhorodecki2005-2} defined by
\begin{equation}\label{partial_info}
\jf{S\parenth{B \mid A}}=S\parenth{\rho_{AB}}-S\parenth{\rho_{A}},
\end{equation}
where \jf{$S(\rho_{AB})$ and $S(\rho_{A})$ stand for the von Neumann entropies of the total
state $\rho_{AB}$ and the  part of the total state held by Alice, $\rho_A$,
respectively}.  The conditional
entropy quantifies the amount of quantum information Bob must
send to Alice such that she can recreate the total state, $\rho_{AB}$, given
her prior knowledge of it, as quantified by $S\parenth{\rho_{A}}$.
As such, the conditional entropy quantifies Alice's ignorance of the
total state.  The coherent information,
Eq.~\parenth{\ref{coherent_info}}, depends on both the channel, $T$,
as well as on the input state, $\rho$, to the channel.  In order \jf{to
evaluate} the quantum capacity of an
arbitrary channel, $T$, the coherent information must \jf{be maximized} over 
all possible input states and regularized  over many uses of the channel. 
For teleporation channels, where $T$ would correspond to a shared 
entangled state \jf{with CM} $\gamma$, however, 
a lower bound to the quantum capacity can be
obtained by first applying a distillation protocol to the \jf{state in order} to obtain $k$ maximally entangled pairs of quantum
bits (ebits).  The teleportation protocol could then be conducted using these ebits.  It was shown
by Wolf \emph{et al}.~\cite{wolf2006} that \jf{the} number of ebits that can be
obtained from a given state \jf{with CM} $\gamma$ \jf{can be bounded from
below} by the right hand side (RHS) of 
\begin{equation}\label{lbound_qcap}
Q\parenth{T}\geq S\parenth{\gamma_A}-S\parenth{\gamma} \equiv Q_L,
\end{equation}
which in turn gives a lower bound to the quantum channel capacity.
\jf{Here $S(\gamma)$ denotes the von Neumann entropy of a Gaussian state with CM
$\gamma$.}
The development of entanglement distillation protocols is an active area
of current research.  A major step towards implementation of
entanglement distillation for continuous variables
\jf{\cite{browne2003,eisert2004}}
has been made by the demonstration of squeezed 
state purification \cite{franzen2006,heersink2006} \jf{and subtraction of single
photons from squeezed states \cite{ourjoumtsev2006,nielsen2006,wakui2006}.}
As shown in \cite{fiurasek2006}, the protocol demonstrated in
\cite{franzen2006} is quite general and can be extended in a
straightforward manner to an iterative purification protocol as
well as to entanglement distillation in the presences of non-Gaussian decoherence.    

In addition to the quantum capacity, there are a number
of \jf{other} quantities that will contribute to the characterization of
our teleportation channels.  To begin with, the state condition,
defined by
\begin{equation}\label{state_cond}
\gamma+i\Omega\geq 0,
\end{equation}
where
\jf{$
\Omega$
is again the two-mode} symplectic form, determines whether the reconstructed CM
corresponds to a physical state \cite{simon2000}.  
\jf{We define $\lambda$ as the minimum eigenvalue of $\gamma+i\Omega$ and 
the inequality (\ref{state_cond}) holds iff $\lambda \geq 0$.}
In order to verify that the channel
has been established using entanglement, the Simon-Peres-Horodecki
(Simon) criterion \cite{simon2000} can be used and can be formulated as
\begin{equation}\label{ppt}
\gamma^{T_{A}}+i\Omega\geq0,
\end{equation}
where $\gamma^{T_{A}}=\Lambda \gamma \Lambda$ is the \jf{CM of a state 
partially transposed with respect to Alice's mode} and
\jf{$\Lambda=\mathrm{diag}\parenth{1,-1,1,1}$} corresponds to a local
time reversal operation on \jf{Alice's} phase quadrature only. 
\jf{Similarly as before, we define $\lambda^{T_A}$ as minimum eigenvalue of
$\gamma^{T_A}+i\Omega$. If $\lambda^{T_A}<0$ then the state is entangled.}
 In addition to the Simon criterion, an optimal entanglement witness,
$\mathcal{W}$, was \jf{determined} by solving the corresponding semi-definite
program \cite{hyllus2006}.  The amount of entanglement was quantified
using the logarithmic \jf{negativity} \cite{vidal2001}, defined by
\begin{equation}\label{logneg}
\logneg=\log_2 \parallel \rho^{T_{A}} \parallel,
\end{equation}       
where a basis 2 sets the units to bits.  The teleporation fidelity for
coherent states \cite{fiurasek2002} is given by
\begin{equation}\label{fidel}
\mathcal{F}=\frac{2}{\sqrt{\mathrm{det}{\bf E}}},
\end{equation}
where the matrix $\bf{E}$ reads
\begin{equation}
{\bf E}=2{\bf D}+{\bf RA}{\bf R}^{T}+{\bf RC}+{\bf C}^{T}{\bf
  R}^{T}+{\bf B},
\end{equation}
and the matrices ${\bf A}$, ${\bf B}$, ${\bf C}$ and ${\bf C}^T$ are obtained
from the CM given by Eq.~\parenth{\ref{cvmat}} with 
\begin{equation}
{\bf R}=\parenth{\begin{array}{cc}
1&0\\
0&-1\end{array}}.
\end{equation}
The purity of the state is defined by \jf{$\mu=\mathrm{Tr}[\rho_{AB}^2]$ and 
for Gaussian states we have}
\begin{equation}\label{purity}
\mu=\frac{1}{\sqrt{\mathrm{det}\gamma}}.
\end{equation}
\jf{Finally we evaluate the achievable secret key rate for entangled
state-based quantum key distribution protocol where Alice and Bob both measure
certain quadrature using local homodyne detections on their parts of the shared
two-mode state. From the knowledge of the covariance matrix $\gamma$ a 
lower bound on the achievable secret key rate can be calcuated by assuming that
the state is Gaussian  and using the following formula,   
\begin{equation}\label{sec_key}
K=I_{AB}-\chi(A:E).
\end{equation}
Here $I_{AB}$ is the classical mutual information between Alice's and Bob's
measured data and $\chi(A:E)$ denotes the Holevo bound between Alice and an
eavesdropper Eve \cite{garcia2006,navascues2006}. This latter quantity can be expressed as
$\chi(A:E)=S(\rho_{AB})-S(\rho_{B}^{a})$, where $\rho_{B}^{a}$ is a
normalized density matrix of Bob's mode conditional on Alice's measurement outcome
$a$. Note that for Gaussian states and homodyne detection $S(\rho_{B}^{a})$
does not depend on the measurement outcome $a$ which justifies the use of the
above expression. }

\begin{figure}
\subfigure[Quadratures]{\includegraphics[width = 9.0cm]{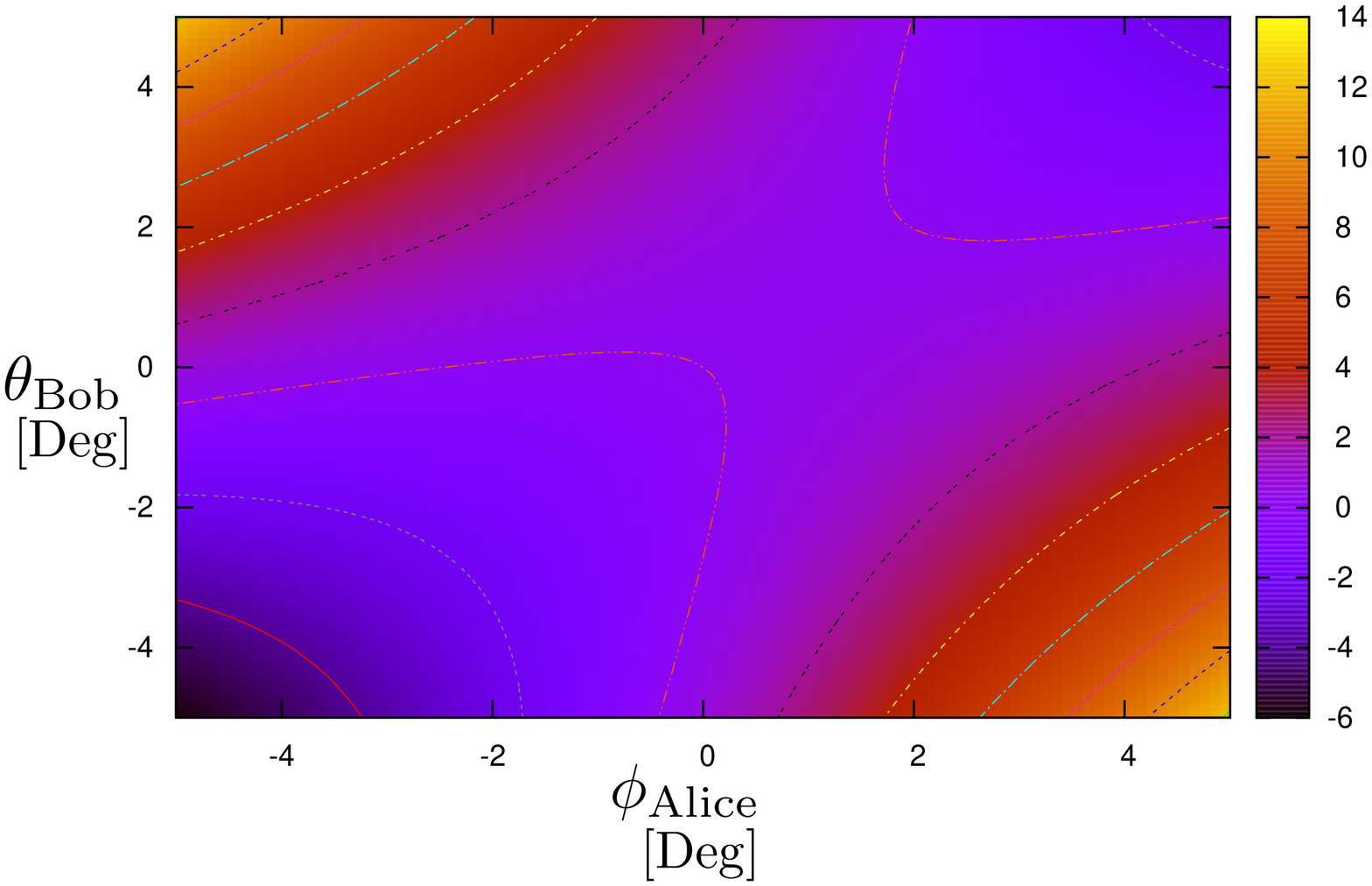}}
\subfigure[Quadrature linear combination]{\includegraphics[width = 9.0cm]{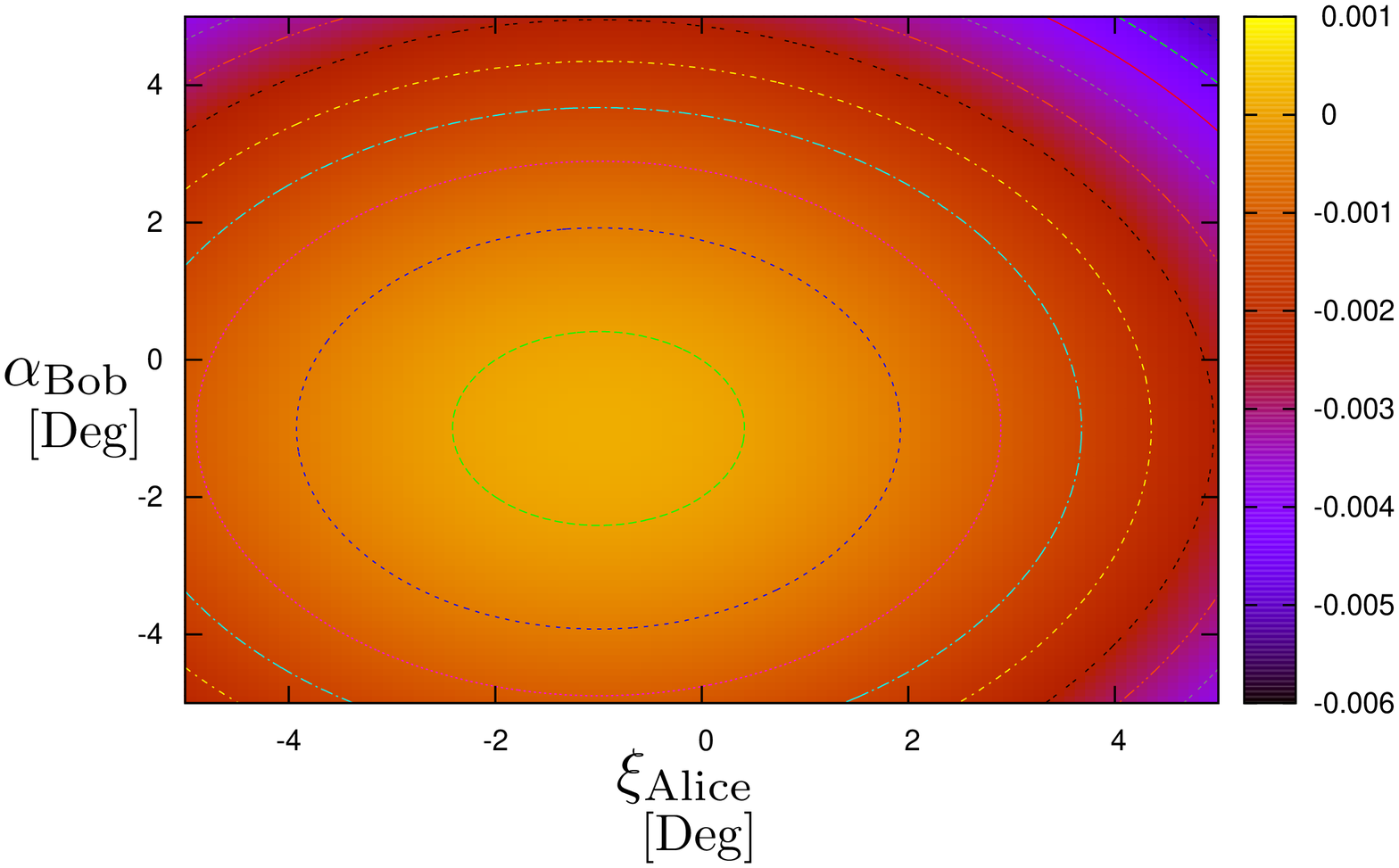}}
\caption{(Color online) Error analysis:  These plots depicted the
  error on the log-negativity for systematic phase offsets on the
  quadratures, Fig.~\ref{error}\parenth{\mathrm{a}}, and on their
  linear combination, Fig.~\ref{error}\parenth{\mathrm{b}}.  It is
  seen that the linear combination parameters are more robust to
  experimental systematic error than the individual quadratures and
  their dependencies.}
\label{error}
\end{figure}

\section{Experimental Results}\label{results}
The partial tomographic protocol (PTP) presented in Sec.~\ref{direc_protocol} was used to
characterize teleportation channels established by two different
classes of distributed bipartite entanglement.  Data acquisition was performed
using a LabView program.  One million data points were recorded per
measurement setting.  The data was then divided into ten separate data blocks
each with 100,000 points. Covariance matrices were generated from each
of the ten and averaged yielding an average CM.  For each CM, the
channel characteristics were calculated and averaged.  The standard
error was then calculated for the $95\%$ confidence interval.  With
respect to the CMs, this ranged from $\pm0.001$ to $\pm0.01$.  The first class, to be known as
V-class entanglement, was formed by mixing a single-mode squeezed vacuum
state with the vacuum mode on a balanced beamsplitter.  According to
the formalism presented by Wolf \textit{et al}.~\cite{wolf2002}, this
represents the optimal entangling scheme for these input states.  This
experiment was conducted for a parametric gain setting of 5 and a
parametric gain setting of 10.  The reconstructed V-Class covariance matrix
(CM) for the parametric gain 5 setting is given by

\begin{center}
\parenth{\begin{array}{cccc}
0.751&-0.146&0.307&-0.000\\
\jf{-0.146}\W&\W3.175&-0.000\W&-2.129\\
0.307&-0.000&0.706&-0.102\\
-0.000\W&-2.129&-0.102\W&\W3.181\\
\end{array}}.
\end{center}

The channel characteristics are presented in Table \ref{sum_results}.  They
include in order of appearance: the state condition Eq.~\parenth{\ref{state_cond}},
the Simon criterion Eq.~\parenth{\ref{ppt}}, an optimal witness; the logarithmic
negativity Eq.~\parenth{\ref{logneg}}, the lower bound to the quantum
channel capacity Eq.~\parenth{\ref{lbound_qcap}}, the teleporation
fidelity of coherent states Eq.~\parenth{\ref{fidel}}, and the purity of the
entangled state Eq.~\parenth{\ref{purity}}.  The
state condition demonstrates that the reconstructed CM is a bona fide
CM i.e., that the CM corresponds to a physical state.  This serves as
an indicator if the measurement has been conducted correctly.  Both
the Simon criterion and entanglement witness serve as a check if the
state is separable or entangled.  The advantage of using an entanglement
witness is that it corresponds to the optimized measuring device that
can be reconstructed from the measured data \cite{hyllus2006}.  As a
result of this optimization, measuring a witness may involve even
fewer measurement settings in order to \jf{optimally detect the entanglement of
the state.} 

\begin{figure}
\begin{center}
\includegraphics[width = 9.0cm]{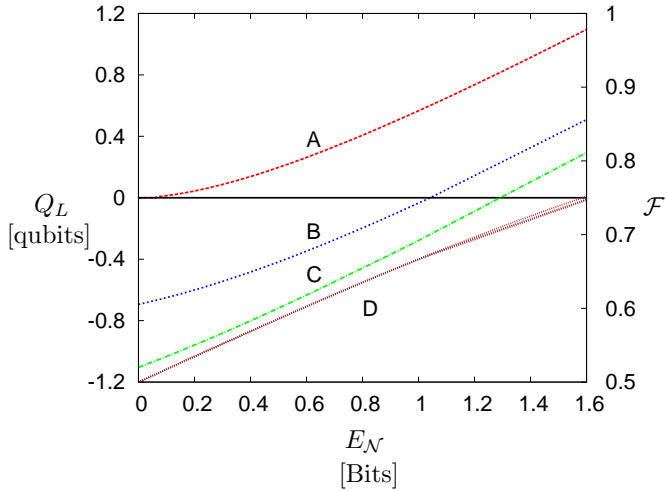}
\caption{(Color online) S-Class entanglement:  This plot depicts the dependence of the
  lower bound to the quantum channel capacity, $Q_L$ on the purity of the entangled state, where A is
  $\mu=1$, B is $\mu=0.5$ and C is $\mu=0.2$, and the
  amount of entanglement.  Interesting for the application of a state
  merging protocol or to obtain a positive secret key rate, is the
  point at which $Q_L$ becomes positive.  It is seen that as the
  purity of the state decreases, more entanglement is required for it
  to become positive.  Curve D is the fidelity for each case.}
\label{sym}
\end{center}
\end{figure}

The V-class parametric gain 10 CM reads

\begin{center}
\parenth{\begin{array}{cccc}
0.686&-0.054&0.326&0.003\\
-0.054\W&\W4.625&0.001&-3.584\W\\
0.326&\W0.001&0.678&-0.031\W\\
0.003&-3.584&-0.031\W&4.681
\end{array}},
\end{center} 
with the corresponding channel characteristics also given in Table
\ref{sum_results}.  In both cases, the lower bound to the channel
capacity is negative.  The teleporation fidelities,
$\mathcal{F}$, both being greater than 1/2, indicate the presence of
entanglement.  The negative values for the Simon criterion and
entanglement witness clearly show the measured state was entangled.  

\begin{figure*}
\subfigure[V-Class entanglement]{\includegraphics[width = 7.0cm]{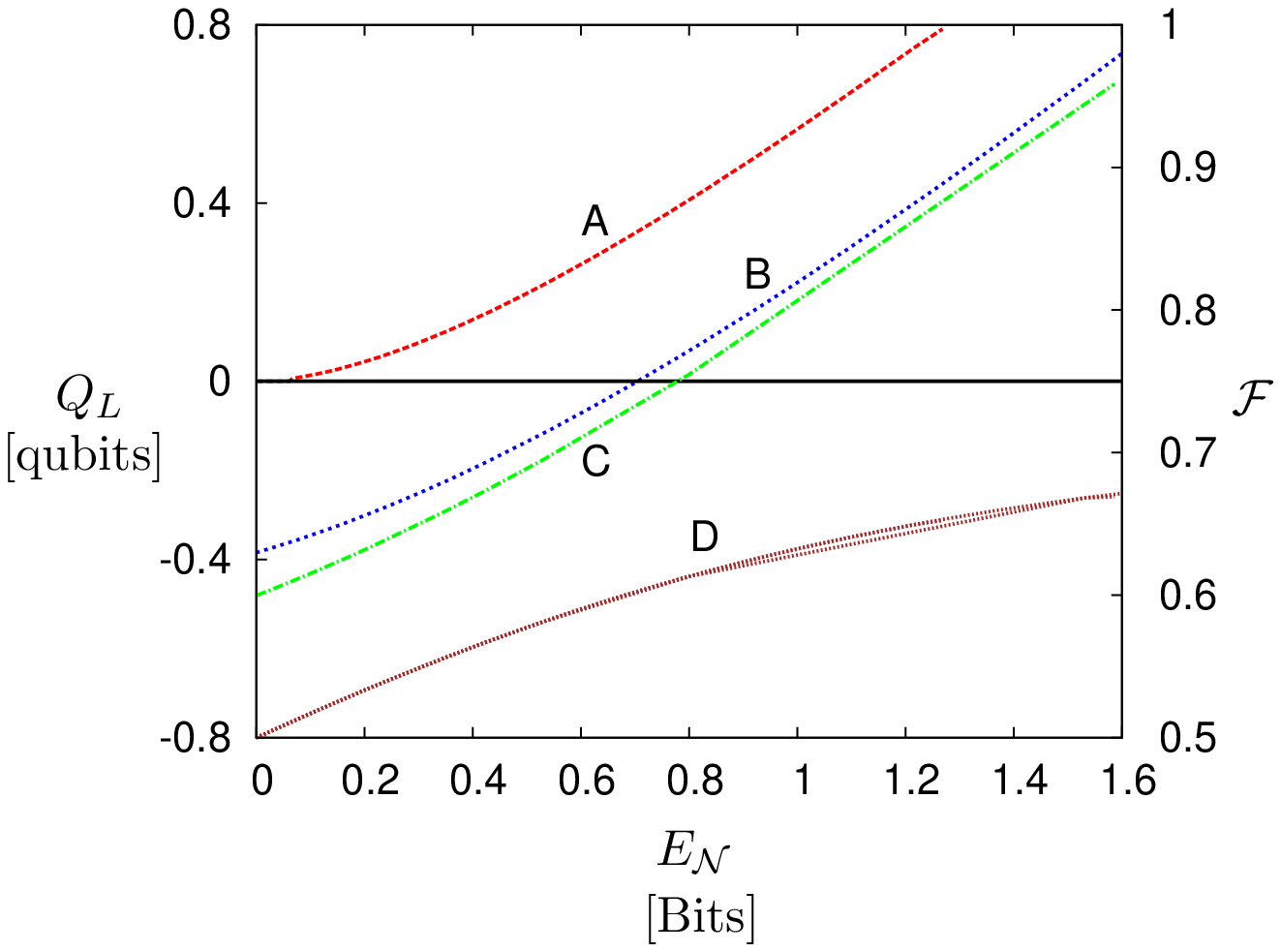}}
\vspace{0.2cm}
\subfigure[M-Class entanglement]{\includegraphics[width = 7.0cm]{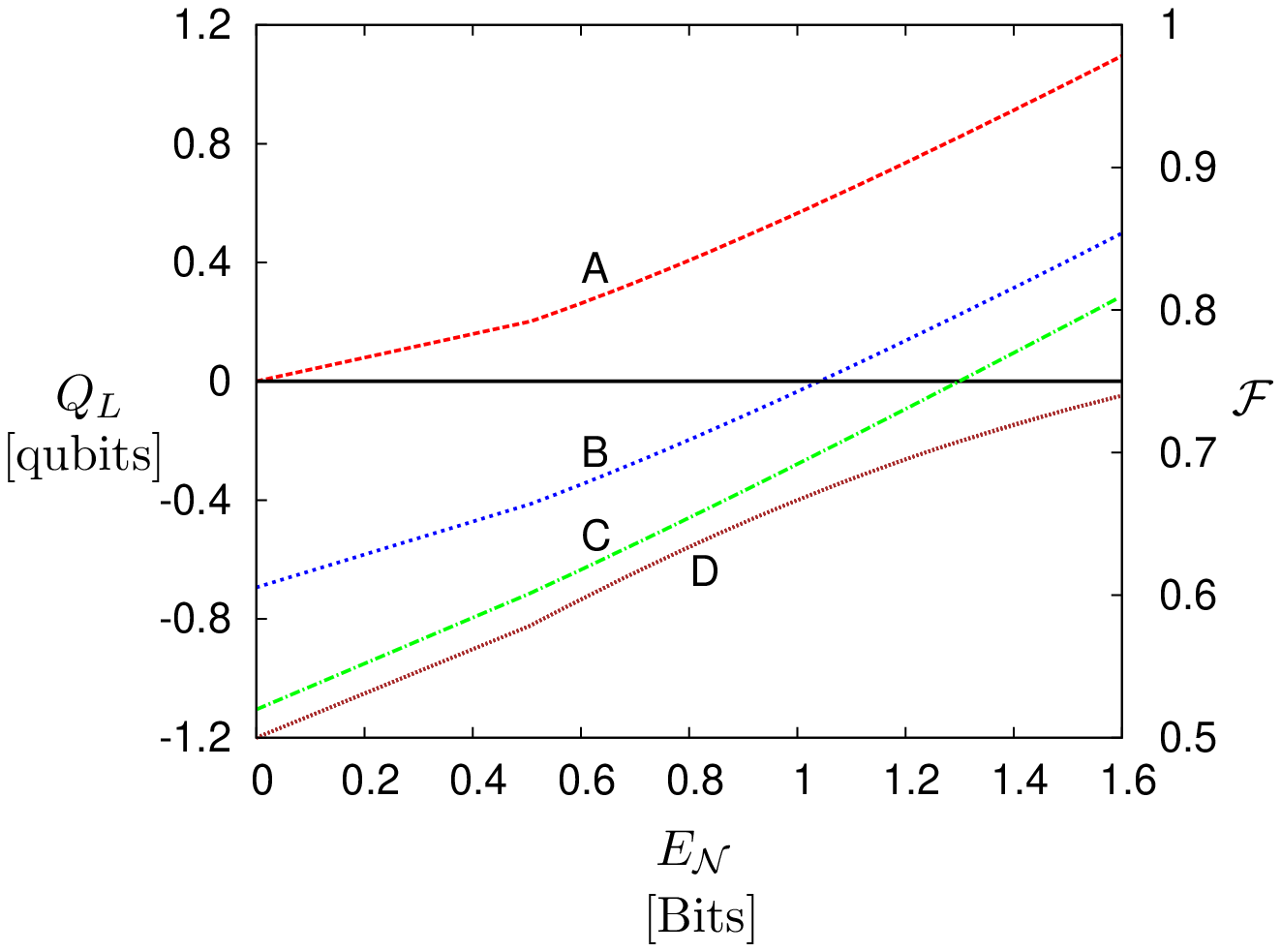}}
\caption{(Color online) V-Class and M-Class entanglement:  These plots were
  created in a similar fashion as Fig.~\ref{sym}, but for the
  V-Class and M-Class entanglement.  As in the previous plot, A
  corresponds to a pure state, $\mu=1$, B to $\mu=0.5$ and C to
  $\mu=0.2$.  In both plots, the partial
  information becomes positive before the S-Class entanglement for the same
  purity.}
\label{nonsymmix}
\end{figure*}

The second class of entanglement generated, to be known as S-Class
entanglement, was established by mixing two equally but oppositely squeezed beams on a
balanced beamsplitter. \jf{For ideal pure squeezed states this would yield the
two-mode squeezed vacuum state.} The reconstructed S-Class CMs for the parametric gain
5 and 10 settings read
\begin{center}
\parenth{\begin{array}{cccc}
2.359&0.132&1.885&0.028\\
0.132&2.205&0.008&-1.883\W\\
1.885&0.008&2.266&0.372\\
0.028&-1.883\W&0.372&2.427
\end{array}},
\end{center}
and
\begin{center}
\parenth{\begin{array}{cccc}
4.200&-0.090&3.773&-0.033\\
-0.090\W&\W4.462&0.035&-4.216\\
3.773&\W0.035&4.228&-0.208\\
-0.033\W&-4.216&-0.208\W&\W4.842
\end{array}},
\end{center}
respectively.  The corresponding channel characteristics are listed in
the third and fourth columns of Table \ref{sum_results},
respectively.  Comparing the two log-negativities of the V-Class and
S-Class entangled states, it is seen that the introduction of another
squeezed beam increases the amount of entanglement for the same
gain setting \jf{roughly by} a factor of 2.  The lower bound to the
quantum capacity also now shows a positive value for each S-Class state.  
The sign \jf{of $Q_L$} is
dependent on both the purity of the state as well as on
the amount of entanglement, a relationship that will be explored more
fully in Sec.\ref{discussion}. The fidelities are both greater than
0.5, indicating the presence of entanglement.  The fidelity of the gain
5 S-Class state, $\mathcal{F}=0.701$, breaks the 2/3 no-cloning limit,
which is experimentally significant \cite{grosshans2001,bowenieee2003}.  The difference of purities can be understood when
considering that for V-Class entanglement the vacuum mode introduces
only a fixed amount of noise whereas for S-Class entanglement, the
extra noise introduced into the entangled state in the form of
anti-squeezing is not fundamentally bounded.

As a final result, a positive secret key rate was obtained from both
S-Class states.  With a resolution bandwidth of 50kHz, 16.1kbit/s of a
secure key could be extracted using the channel established by the
parametric gain 5 setting. \jf{This shows 
that the S-Class entangled states can be used for 
continuous variable quantum cryptography.} 
Additionally, drawing on the recent results of Horodecki \textit{et
  al}.~\cite{mhorodecki2005,mhorodecki2005-2}, the positive $Q_L$'s
indicate that \textit{state merging} can be achieved using only local
operations and classical communication.       

\section{Discussion}\label{discussion}
It is seen that non-zero entries appear \jf{in reconstructed CMs} 
where zero normally would have
been expected.  The question remains at what point do these non-zero
entries become significant?  An analysis of the error incurred as a
result of either falsely measuring, or not measuring at all, the
non-standard entries has been conducted in \cite{laurat2005} and a
similar analysis is presented here.  In Fig.~\ref{error} is illustrated
the percent error on the log-negativity in the presence of
phase offsets on the homodyne detectors.  Two cases are numerically
investigated: Fig.~\ref{error}\parenth{\mathrm{a}} is the effect of phase
offsets on measuring the amplitude and phase quadratures.  The effect
of phase offsets on measuring the linear combination of the quadratures assuming the individual
quadratures have been properly measured is illustrated in
Fig.~\ref{error}\parenth{\mathrm{b}}.  The simulation was conducted
by generating a covariance matrix corresponding to a pure
non-optimally entangled \jf{S-class state}, whose quadrature
variances were dependent on four independent parameters, namely;
$\phi_{\mathrm{Alice}}$\,,$\theta_{\mathrm{Bob}}$\,,$\xi_{\mathrm{Alice}}$ and $\alpha_{\mathrm{Bob}}$.  The
parameters $\phi$ and $\theta$ correspond to the independent phase
offsets applied to Alice's and Bob's quadratures, respectively.
The parameters $\xi$ and $\alpha$ correspond to the independent phase
offsets applied to Alice's and Bob's linear combination of
quadratures.  In this way, the effect of incorrectly measuring the
quadratures on the value of the covariance between them can be
analyzed.  In Fig.~\ref{error}\parenth{\mathrm{a}} the percent error on the
log-negativity dependent on phase offsets on Alice's and Bob's
quadratures is presented.  It is seen that in the region of $\phi \,,
\theta \in \jf{[-2,2]}$, the percent error can be as
high as -2\%, indicating that the amount of entanglement is
underestimated.  As the phase offsets increase, so does the error
reaching as high as 14\% overestimation and -6\% underestimation.  The
percent error dependent on phase offsets on the linear combination
terms is presented in Fig.~\ref{error}\parenth{{b}}.  It is seen that the
error is significantly less for the same region as in the quadrature
case.  This illustrates that the linear combination terms are far more
robust to systematic error than the quadrature terms and their dependencies.    

The experimental results highlight a relationship between the
purity, log-negativity and the lower bound to the quantum channel
capacity $Q_L$.  This relationship is made explicit in the numerical results presented in
Figs.~\ref{sym} and \ref{nonsymmix}.  The $Q_L$
for S-Class entanglement is shown for three different purities in
Fig.~\ref{sym}.  It is seen that for pure states i.e., $\mu=1$, the presence
of entanglement ensures a positive $Q_L$.  As the purity of the
state decreases, the zero crossing is shifted towards higher levels of
entanglement.  The teleportation fidelity, plotted on the second
abscissa, is independent of the purity of the state (assuming that the
channel has been properly homodyned).  The purity dependence of the
$Q_L$ can be further investigated by looking at its behavior for two
other classes of entanglement.  The $Q_L$ for V-Class entanglement and
for M-Class entanglement, formed by mixing two unequally and oppositely squeezed beams on a
balanced beamsplitter, is shown in Fig.~\ref{nonsymmix}.  It is readily seen that the zero crossing for
less than pure states occurs at lower levels of entanglement than for
S-Class entanglement.  Indeed, the numerical results for M-Class
entanglement manifest optimal behavior for less than pure
states, being positive earlier than for S-Class entanglement for the same
purity.  Although all three entangled states are bipartite Gaussian
states, their utility is very much dependent on their underlying construction.

\section{Conclusion}
In this work we have presented an efficient method for the
characterization of Gaussian communication channels with which the
\textit{entire covariance matrix} was measured.  This method was
applied to two different classes of continuous variable entangled
states which were used to establish a teleportation channel between
distant parties.  The lower bound to the quantum channel capacity as well as
other characteristics of the channel were evaluated from the
reconstructed covariance matrix.  The relationship between the purity,
entanglement class, and quantum channel capacity were explored
numerically.  Two of the established teleportation channels delivered
both a positive $Q_L$ as well as a positive secret key rate.

\begin{acknowledgments}
We acknowledge financial support from the Deutsche
Forschungsgemeinschaft (DFG), project number SCHN 757/2-1.  J.~F.~
acknowledges financial support from the Ministry of Education
of the Czech Republic under the projects Centre for Modern Optics
(LC06007) and Measurement and Information in Optics (MSM6198959213)
and from the EU under project COVAQIAL (FP6-511004).  We also thank Dr.~Paul
Cochrane for fruitful conversations and for reading the manuscript
prior to publication.
\end{acknowledgments}

% Produces the bibliography via BibTeX.

%\bibliography{purification,sqz,teleport,entmeas,qchannels,qinformation,entgeneral,entcriterions,qkd,nonclassical_states,mathphys}

\end{document}